\documentclass[ aps,%
floatfix,%
final,%
notitlepage,%
oneside,%
onecolumn,%
nobibnotes,%
nofootinbib,%
superscriptaddress,%
showpacs,%
]%
{revtex4}

\usepackage{graphicx}

\begin{document}

\title{Scalar mesons in $\eta'\to3\pi^{0},\pi^{0}\pi^{+}\pi^{-}$ decays
}
\author{A. K. Likhoded}
\email{Anatolii.Likhoded@ihep.ru}
\author{A. V. Luchinsky}
\email{Alexey.Luchinsky@ihep.ru}
\author{V. D. Samoylenko}
\affiliation{Institute for High Energy Physics, Protvino, Russia}

\begin{abstract} 
The decays $\eta'\to3\pi^{0}$ and $\pi^{0}\pi^{+}\pi^{-}$ were consider in isobar model approach. Branching ratio and Dalitz-plot shape of the decay $\eta'\to3\pi^{0}$ can be explain by $a0$- and $\sigma$-meson contributions. The $\sigma$ meson term is needed for the Dalitz-plot distribution reconstruction. The branching ratio of the decay $\eta'\to \pi^{0}\pi^{+}\pi^{-}$ can be calculated in the same isobar framework. The prediction for the Dalitz-plot of this decay is shown which strongly dependents from model parameters.
\end{abstract}

\maketitle 

\section{Introduction}

Light scalar mesons are the subject of extensive theoretical and experimental investigation in recent years (see. \cite{Narison2008nj} and references therein). The structure of these mesons in terms of constituent quarks and gluons in rather vague. There are several models for light scalar mesons: usual quark-antiquark states, tetraquark, meson-meson molecules, etc. Glueball states are also expected in $\approx 1$ GeV mass region scalar sector.

The problem of light scalar mesons studying is that some of these states ($\sigma$- and $\kappa$-mesons) have large widths comparable with their mass, while the other ($a_{0}(980)$ and $f_{0}(980)$) lie near inelastic thresholds. For this reason one cannot expect usual Breit-Wigner form of the signal. There is also one class of reactions (for example $\eta'\to\eta\pi\pi$ decay), that require the contributions of scalar mesons. Usually such reactions are considered in the framework of Chiral Perturbation theory \cite{Tornqvist1995kr,Majumdar1968aa,Cronin1967jq, Schwinger1968zz,DiVecchia1980sq,Beisert2002ad,Bijnens2005sj}). although for large energies the use of this formalism could be unacceptable.

We think, that there is the other, more physically motivated approach to such decays. In the framework of so called isobar model the matrix element is saturated by contributions of virtual mesons with suitable masses and quantum numbers. In a series of works (\cite{Deshpande1978iv,Fariborz1999gr, Borasoy2005du,Donskov2009ri}) it was shown, that in order to describe the decay $\eta'\to\eta\pi\pi$ it is sufficient to take into account the contributions of $\sigma$- and $a_0$-mesons. The branching fraction of this decay is reproduced by $a_0$-meson contribution, and in order to explain the form of the Dalitz-plot, the contribution of $\sigma$-meson should also be taken into consideration.

In our article we use isobar model to study $\eta'\to3\pi^{0}$, $\pi^{0}\pi^{+}\pi^{-}$ decays. It is well known that in these reactions $G$-parity is violated. According to Sutherland theorem \cite{Sutherland1966zz} the electromagnetic interaction in this case can be neglected, so only the difference of $u$- and $d$-quark masses can give the isospin violation. For this reason the study of $\eta'\to 3\pi$ decays could give valuable information on the ratios of current quark masses. In the framework of isobar model isospin violation is naturally parametrized in the terms of mixing of states with different isospin values and small mixing parameters should be proportional to $m_d-m_u$ mass difference. In our article we determine these parameters from fit of experimental branching fractions and Dalitz-plot distributions of $\eta'\to3\pi^{0}$, $\pi^{0}\pi^{+}\pi^{-}$ decays.

The rest of the paper is organized as follows. In the next section we describe in detail the mechanism of isospin violation in the framework of isobar model and present diagrams that give contribution to $\eta'\to3\pi^{0}$, $\pi^{0}\pi^{+}\pi^{-}$ decays. In sec. III and IV we use obtained matrix elements to analyze neutral and charged modes of $\eta'\to 3\pi$ decays. Short discussion is presented in Conclusion. In the Appendix we give the explicit form of used in our paper $\sigma$-, $a_0$-, and $\rho$-meson exchange amplitudes.

\section{Models of isospin violation}

Let us consider isospin violation decays $\eta'\to 3\pi$ in the framework of isobar model, when the amplitude of the process is saturated by the contributions of virtual mesons with suitable masses and quantum numbers.

This approach was used earlier to describe the isospin-conserving decay $\eta'\to\eta\pi^{0}\pi^{0}$ \cite{Deshpande1978iv,Fariborz1999gr,
Borasoy2005du,Donskov2009ri}. In this works it was shown, that in order to reproduce the form of experimental Dalitz-plot it is sufficient to take into account the contributions of two scalar resonances: $\sigma$-meson in $\pi^0\pi^0$-channel and $a_0$-meson in $\pi^0\eta$-channel. The matrix element of this decay can be written in the form
\begin{eqnarray*}
\mathcal{A}\mbox{\ensuremath{\left(\eta'\to\eta\pi^{0}\pi^{0}\right)}} & =
& \mathcal{A}_{\sigma}\left(s_{\pi\pi}\right)+\mathcal{A}_{a}
\left(s_{\eta\pi_{1}}\right)+\mathcal{A}_{a}\left(s_{\eta\pi_{2}}\right),
\end{eqnarray*}
where $s_{ij}$ are invariant masses of corresponding pairs squared and $\mathcal{A}_{\sigma}(s),\mathcal{A}_{a}(s)$ are the rescattering amplitudes in $\pi\pi$- and $\pi\eta$-channels (explicit expressions for these amplitudes can be found in the Appendix). In our analysis of $\eta'\to3\pi$ decays we also restrict ourselves to contributions of these two resonances.

When describing the isospin violation in considered here decays it is necessary to recall, that physically observed mesons have no definite value of isospin $I$. For example, $\pi^0$- and $\eta$-mesons are actually superposition of $I=0$ and $I=1$ states
\begin{eqnarray*}
\left|\pi^{0}\right\rangle _{\mbox{phys}} & = 
& \left|\pi^{0}\right\rangle _{I=1}+\epsilon\left|\pi^{0}\right\rangle _{I=0},
\qquad\left|\eta\right\rangle _{\mbox{phys}}=\left|\eta\right\rangle _{I=0}+
\epsilon\left|\eta\right\rangle _{I=1}.
\end{eqnarray*}
The wave functions of other involved in the reaction particles can also be written in a similar form. The admixtures with ``wrong'' isospin values ($\left|\pi^0\right\rangle_{I=0}$ and $\left|\eta\right\rangle_{I=1}$ in the presented above examples) should be suppressed by the small factor
\begin{eqnarray*}
\epsilon\sim\sin\lambda & \sim & \frac{m_{d}-m_{u}}{m_{s}}\sim10^{-2},
\end{eqnarray*}
that is usually interpreted as the mixing angle between states with different isospin values \cite{Gross1979ur}. In what follows we will consider only linear in this parameter terms.

\begin{figure}
\includegraphics{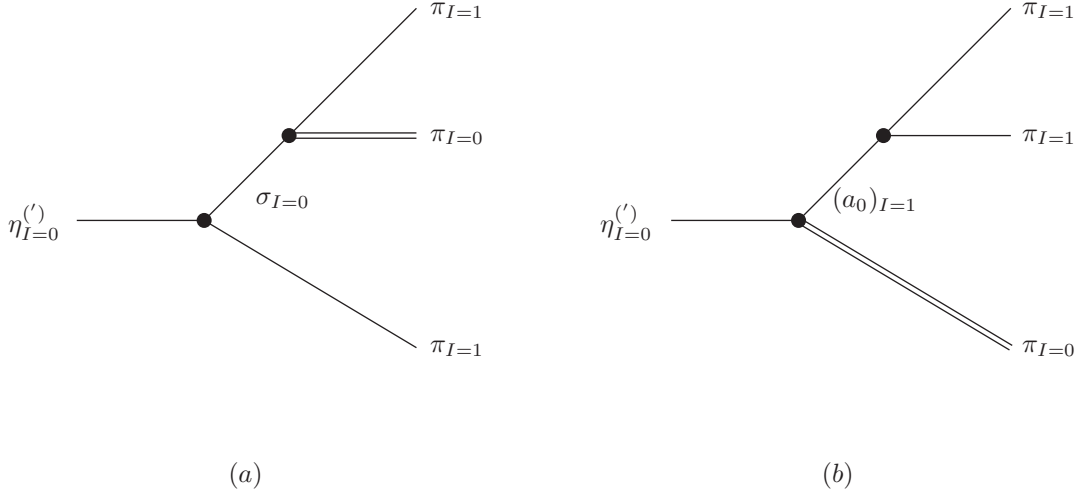}
\caption{Doubly isospin violated diagrams of $\eta'\to3\pi$ decay. The particles shown with double line have isospin different from tabular value
}
\label{diag:forbidden} 
\end{figure}

This restriction reduces significantly the number of diagrams that can give contribution to considered here processes. For, example, one can neglect the diagram shown in fig.~\ref{diag:forbidden}a, since in the decay $\left|\sigma\right\rangle _{I=0}\to\left|\pi^{0}\right\rangle _{I=1}\left|\pi^{0}\right\rangle _{I=0}$ isospin is violated. As a result the corresponding diagram is quadratic in $\epsilon$ and it can be neglected. The same is valid also for the diagram shown in fig.~\ref{diag:forbidden}b,  where additional suppression factor stems from generalized Bose-Einstein symmetry, that in the isospin-conservation limit forbids the situation when two $\pi$-mesons are in the states with zero orbital momentum and total isospin $I_{\pi\pi}=1$. For this reason in the vertex $\left|a_{0}\right\rangle _{I=1}\to\left|\pi^{0}\right\rangle _{I=1}\left|\pi^{0}\right\rangle _{I=1}$ isospin is violated.

\begin{figure}
\includegraphics{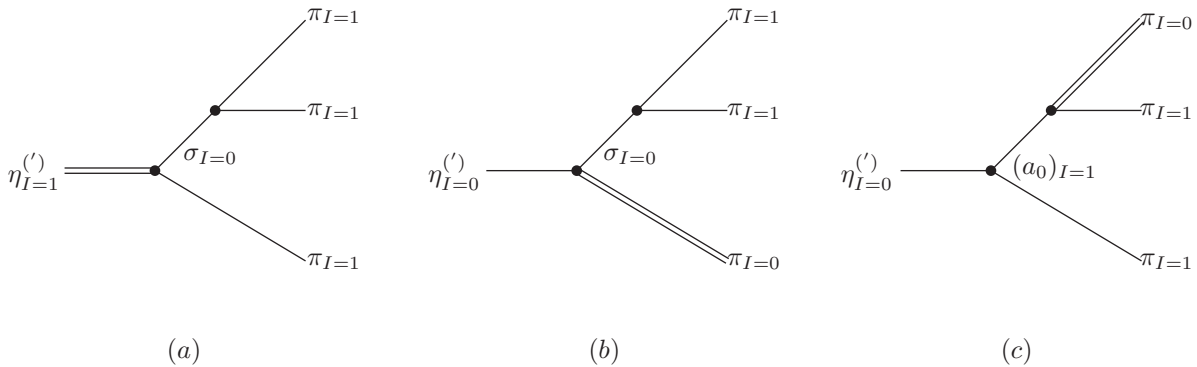}
\caption{
Diagrams of $\eta'\to3\pi$ decays, that violates isospin only once. The particles shown with double line have isospin different from tabular value
}
\label{diiag:Allowed} 
\end{figure}

Using these arguments it is easy to see, that only diagrams shown in fig.~\ref{diiag:Allowed} can give linear in isospin violation parameter contributions to the considered in our article processes. The explicit form of the matrix element depends on what final state ($3\pi^0$ or $\pi^0\pi^+\pi^-$) we are talking about. In the case of neutral final state 
\begin{eqnarray}
\eta^{'} & \to & \pi^{0}(p_{1})\pi^{0}(p_{2})\pi^{0}(p_{3}),
\label{etaP3Pi}
\end{eqnarray}
the amplitude should be symmetric with respect to momenta of all final particles. In is clear, that it can be written in the form
\begin{eqnarray}
\mathcal{A}\left[3\pi^{0}\right] & = & \epsilon_{\sigma}\mathcal{A}\left(s_{1}\right)+
\epsilon_{a}\mathcal{A}_{a}\left(s_{1}\right)+\left(s_{1}\to 
s_{2}\right)+\left(s_{1}\to s_{3}\right),
\label{amp:neutral}
\end{eqnarray}
where $s_{ij}=\left(p_{i}+p_{j}\right)^{2}$, amplitudes $\mathcal{A}_{a,\sigma}(s)$ are given in the Appendix, and small parameters $\epsilon_{a,\sigma}$ will be determined from the fit of experimental data. In the case of charge final state
\begin{eqnarray}
\eta^{'} & \to & \pi^{+}\left(p_{1}\right)\pi^{-}\left(p_{2}\right)\pi^{0}\left(p_{3}\right)
\label{dec:charged}
\end{eqnarray}
no additional symmetrization is required. It is evident, that the particle with $I=0$ (e.g. $\sigma$-meson on diagrams \ref{diiag:Allowed}a,b or the upper of $\pi$-mesons on diagram \ref{diiag:Allowed}c) is neutral. So the amplitude of the decay (\ref{dec:charged}) has the form
\begin{eqnarray}
\mathcal{A}\left[\pi^{0}\pi^{+}\pi^{-}\right] & = & \epsilon_{\sigma}\mathcal{A}_{\sigma}\left(s_{12}\right)
+\epsilon_{a}\left\{ \mathcal{A}_{a}\left(s_{13}\right)+\mathcal{A}_{a}\left(s_{23}\right)\right\}.
\label{amp:charged}
\end{eqnarray}

In some works (see \cite{Borasoy2006uv}) $\rho$-meson contribution to these processes is also considered. It is clear, that the amplitude of $=eta'\to3\pi^0$ decay cannot contain contribution from $\rho$-meson exchange, but in the case of charged final state this is not the fact. So, we should add a term $\epsilon_{\rho}\mathcal{A}_{\rho}\left(s_{12},s_{13}\right)$ to the amplitude (\ref{amp:charged}). The explicit expression for $\mathcal{A}_{\rho}$ is given in the Appendix.

\section{$\eta'\to3\pi^{0}$}

Let us first consider the neutral decay mode $\eta'\to3\pi^{0}$. We will approximate the matrix element of this decay by the following expression
\begin{eqnarray}
\mathcal{A}\left(\eta'\to3\pi^{0}\right) & = & \epsilon_{\sigma}\left\{
\mathcal{A}_{\sigma}\left(s_{12}\right)+\mathcal{A}_{\sigma}\left(s_{13}\right)+\mathcal{A}_{\sigma}\left(s_{23}\right)\right\}
+\epsilon_{a}\left\{
\mathcal{A}_{a}\left(s_{12}\right)+\mathcal{A}_{a}\left(s_{13}\right)+\mathcal{A}_{a}\left(s_{23}\right)\right\}.
\label{eq:Amp3Pi0}
\end{eqnarray}
Here $s_{ij}=\left(p_{i}+p_{j}\right)^{2}$, $\mathcal{A}_{\sigma}(s)$, $\mathcal{A}_{a}(s)$ are amplitudes of $\sigma$- and $a_0$-meson exchange (see Appendix for details), and $\epsilon_{a,\sigma}$ are small isospin violation parameters that will be determined from experimental information about the branching fraction and dalitz-plot shape of this decay.

The branching fraction of the decay $\eta'\to3\pi^{0}$ is equal to \cite{Blik2008zz} 
\begin{eqnarray}
\mbox{BR}_{exp}\left(\eta'\to3\pi^{0}\right) & = & \left(1.61\pm0.23\right)\times10^{-3}.
\label{eq:Br1}
\end{eqnarray}
The dependence of this branching fractions on the model parameters can be written in the form
\begin{eqnarray}
\mbox{BR}_{th} & = & \epsilon_{\sigma}^{2}B_{\sigma\sigma}+\epsilon_{a}^{2}B_{aa}+
2\epsilon_{\sigma}\epsilon_{a}B_{\sigma a},
\label{eq:Brth}
\end{eqnarray}
where  $B_{\sigma\sigma}$, $B_{aa}$ and  $B_{\sigma a}$ are branching fractions of the considered here decays with only $\sigma$-meson, $a_0$-meson and $\sigma-a_0$ interference contributions taken into account. Using presented in the Appendix matrix elements we get numerical values of these coefficients, presented in the second column of table \ref{tab:pi0}.

\begin{table*}
\caption{
Contributions of different channels to the branching fraction and slope coefficient for $\eta'\to3\pi^{0}$ decay (see eq.(\ref{eq:Brth}), (\ref{eq:beta}))
\label{tab:pi0}}
\begin{center}
\begin{tabular}{|c|c|c|c|}
\hline 
$ij$  & $B_{ij}$  & $N_{ij}$  & $\beta_{ij}$\tabularnewline
\hline
\hline 
$\sigma\sigma$  & $2.4$  & $110$  & $-0.67$\tabularnewline
\hline 
$\sigma a$  & $1.4$  & $46$  & $-0.43$\tabularnewline
\hline 
$aa$  & $1.1$  & $53$  & $-0.06$\tabularnewline
\hline
\end{tabular}
\end{center}
\end{table*}

\begin{figure}
\includegraphics{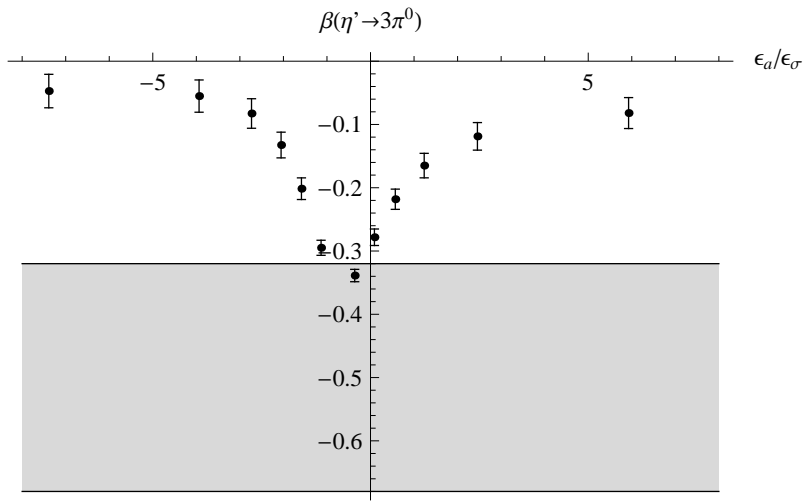}
\caption{Dalitz plot slope coefficient of $\eta'\to3\pi^{0}$ versus the model parameter ratio
}
\label{fig:beta}
\end{figure}

The form of the dalitz-plot is usually written as an expansion in a variable $Z$ defined as
\begin{eqnarray*}
Z & = & \frac{6}{\left(M_{\eta'}-3m_{\pi}\right)^{2}}\sum_{i=1}^{3}\left(T_{i}-
\frac{M_{\eta'}}{3}\right)^{2},
\end{eqnarray*}
Where $T_{1,2,3}$ are kinetic energies of $\pi$-mesons in initial meson rest frame. The parameter $\beta$ in the expansion
\begin{eqnarray}
\left|\mathcal{M}\right|^{2} & = & N\left(1+2\beta Z+\dots\right).
\label{eq:matr0}
\end{eqnarray}
is called the slope coefficient of the dalitz-plot. Experimental value of this coefficient is \cite{Blik2008zz}
\begin{eqnarray}
\beta_{\mbox{exp}}\left(\eta'\to3\pi^{0}\right) & = & -0.59\pm0.18.
\label{eq:beta1}
\end{eqnarray}
The dependence of the slope coefficient on the model parameters can be written in the form
\begin{eqnarray}
\beta_{\mbox{th}} & = & \frac{\epsilon_{\sigma}^{2}N_{\sigma\sigma}
\beta_{\sigma\sigma}+\epsilon_{a}^{2}N_{aa}\beta_{aa}+
2\epsilon_{\sigma}\epsilon_{a}N_{\sigma a}
\beta_{\sigma a}}{\epsilon_{\sigma}^{2}N_{\sigma\sigma}+
\epsilon_{a}^{2}N_{aa}+2\epsilon_{\sigma}\epsilon_{a}N_{\sigma a}},
\label{eq:beta}
\end{eqnarray}
where $N_{\sigma\sigma}$,  $N_{aa}$, $N_{\sigma a}$ and $\beta_{\sigma\sigma}$, $\beta_{aa}$, $\beta_{\sigma a}$ are the parameters of the expansions (\ref{eq:matr0}) of squared matrix elements with only $\sigma$-meson, $a_0$-meson and $\sigma-a_0$-interference contributions taken into account. Numerical values of these parameters are presented in the third and fourth columns of table \ref{tab:pi0}. It is clearly seen, that the slope coefficient of the dalitz-plot with only $a_0$-contribution taken into account (i.e. parameter $\beta_{aa}$) is small. In other words, this resonance gives almost flat distribution over the dalitz plot, so it is necessary to consider also $\sigma$-meson to explain the experimental dependence of the squared matrix element on the invariant masses. Similar situation is observed also for $\eta'\to\eta\pi^{0}\pi^{0}$ decay, where for explanation of experimental dalitz plot $\sigma$-meson play crucial role \cite{Donskov2009ri}. It can also be noted, that the slope coefficient depends only on the ratio $\epsilon_{a}/\epsilon_{\sigma}$. In fig.~\ref{fig:beta} we show the dependence of the slope coefficient on this ratio and demonstrate existing experimental constraints on $\beta$.

\begin{figure} 
\includegraphics{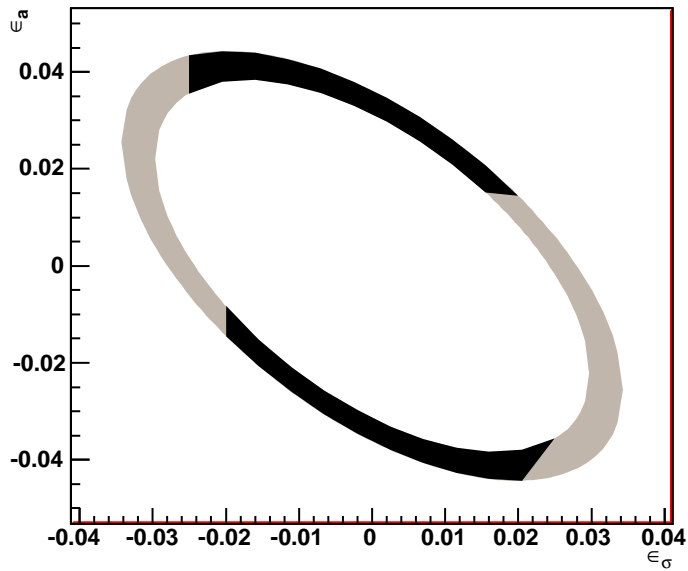}
\caption{
Allowed parameters ($\epsilon_{\sigma}, \epsilon_{a}$) region for $\eta'\to3\pi^{0}$. The widths of the ellipse is equal to $2\sigma$ (the error in branching fraction of this decay). The shaded arcs show the regions with given confidence level.
}
\label{fig:domain1}
\end{figure}

In order to have branching fraction if the decay $\eta'\to3\pi^{0}$ consistent with the experimental value, parameters $\epsilon_{a,\sigma}$ should lie in the region  bounded by two ellipses shown in fig.~\ref{fig:domain1}. The width of this region is determined by the experimental error of the decay (\ref{eq:Br1}). It can be clearly seen, that by an order of magnitude this region coincides with the theoretical estimates presented in \cite{Gross1979ur}. In our analysis we do not restrict ourselves to linear parametrization (\ref{eq:matr0}) and fitted experimental dalitz-plot by the resonance parametrization (\ref{eq:Amp3Pi0}) with fitting parameters $\epsilon_{\sigma,a}$. It turns out the at $\alpha=0.1$ confidence level\footnote{subroutine PROB from \cite{hbook}} there are two symmetric regions on the ellipse (fig.~\ref{fig:domain1}) where required confidence level can be achieved.

These regions are rather large. On the one hand this is because the matrix element of the decay is not very sensitive to the ratio $\epsilon_{\sigma}/ \epsilon_{a}$. On the other hand, the large width of these regions reflects significant experimental errors in slope coefficient of the matrix element, caused by experimental difficulties of the considered decay. For example, in order to remove intensive background signals one has to remove some regions from the dalitz-plot \cite{Blik2008zz}. Moreover, originally the slope coefficient was extracted from experimental data assuming the the squared matrix element is linear in the variable $Z$, and our subsequent analysis shows the presence of higher non-linear terms.

\begin{figure}
\includegraphics{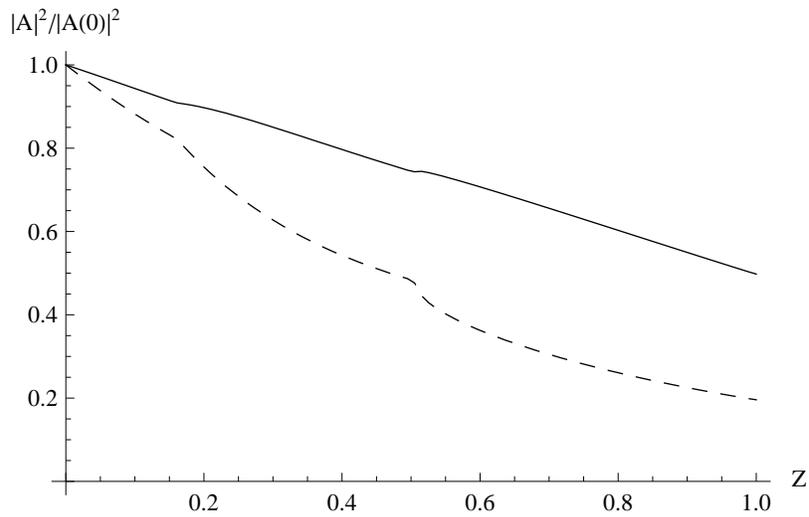}
\caption{
$Z$-distributions for $\eta'\to3\pi^{0}$ decay. Solid and dashed lines correspond to theoretical predictions with parameters $\epsilon_{\sigma}=\epsilon_{a}=0.016$ and  $\epsilon_{\sigma}=0.0220$, $\epsilon_{a}=-0.011$ respectively}
\label{fig:Zdist}
\end{figure}

In fig.~\ref{fig:domain1} the allowed regions of model parameter values are shown by shaded ellipse parts. As it was mentioned above, one can change these parameters without noticeable decrease in experimental data description. In order to demonstrate this fact and to check $Z$-distribution sensitivity to parameter values we considered two cases: equal values of parameters in both channels and the configuration with $ |\epsilon_{\sigma}| > |\epsilon_{a}|$. In the first case one should set
\begin{eqnarray*}
\epsilon_{\sigma} & = & \epsilon_{a}=0.016 \pm0.003,
\end{eqnarray*}
to get the experimental value of the $\eta'\to3\pi^0$ branching fraction. In the second case we have $\epsilon_{\sigma}=0.020 \pm 0.004$, $\epsilon_{a}= - 0.011 \pm 0.002$. In fig.~\ref{fig:Zdist} $Z$-distributions of theoretical predictions of squared matrix element for presented above parameter values are shown. It can be seen, that these curves are different mostly at large $Z$, i.e. at the end of dalitz region, where experimental errors are large. Moreover, as it was mentioned above, in our parametrization the shape of $Z$-distribution depends only on the ratio of the model parameters, for which the uncertainties are compensated with good accuracy.

\section{$\eta'\to\pi^{0}\pi^{+}\pi^{-}$}

Let us now proceed to $\eta'\to\pi^{0}\pi^{+}\pi^{-}$ decay. According to results of sec.II the matrix element of this decay can be written in the form
\begin{eqnarray*}
\mathcal{A}\left[\eta'\to\pi^{0}\left(p_{1}\right)\pi^{+}\left(p_{2}\right)\pi^{-}\left(p_{3}\right)\right] & =
& \epsilon_{\sigma}\mathcal{A}_{\sigma}\left(s_{23}\right)+\epsilon_{a}\left\{ \mathcal{A}_{a}\left(s_{12}\right)
+\mathcal{A}_{a}\left(s_{13}\right)\right\} +\epsilon_{\rho}\mathcal{A}_{\rho}\left(s_{12},s_{13}\right),
\end{eqnarray*}
where $\mathcal{A}_{\sigma,a,\rho}$ are the amplitudes of virtual $\sigma$-, $a_{0}$- and  $\rho$-mesons exchange and $\epsilon_{\sigma,a,\rho}$ are isospin violation parameters ion these channels. Numerical values for $\epsilon_{\sigma,a}$ were obtained in the previous section from $\eta'\to3\pi^{0}$ decay analysis, and $\epsilon_\rho$ can be determined from the fit of experimental data.

\begin{table*}
\caption{Coefficients $B_{ij}$ for $\eta'\to\pi^{0}\pi^{+}\pi^{-}$ decay}
\begin{center}
\begin{tabular}{|c|c|c|c|} \hline 
$i/j$  & $\sigma$  & $a_{0}$  & $\rho$\tabularnewline
\hline
\hline 
$\sigma$  & $1.8$  & $1.4$  & $4.4$\tabularnewline
\hline 
$a_{0}$  & $4.4$  & $3.9$  & 0\tabularnewline
\hline 
$\rho$  & $4.4$  & 0  & $5\times10^{2}$\tabularnewline
\hline
\end{tabular}
\end{center}
\label{tab:charged}
\end{table*}

In the case of $\eta'\to\pi^{0}\pi^{+}\pi^{-}$ only the branching fraction is known experimentally \cite{2008tb}:
\begin{eqnarray*}
BR_{exp}\left(\eta'\to\pi^{0}\pi^{+}\pi^{-}\right) & =
& \left(37_{-9}^{+11}\pm4\right)\times10^{-4}.
\end{eqnarray*}
It is convenient to write model parameter dependence of the theoretical prediction of this branching fraction in the form similar to expression (\ref{eq:Brth}):
\begin{eqnarray*}
BR_{th}\left(\eta'\to\pi^{0}\pi^{+}\pi^{-}\right) & = 
& \sum_{i,j=\sigma,a,\rho}\epsilon_{i}\epsilon_{j}B_{ij}.
\end{eqnarray*}
Numerical values of the coefficients $B_{ij}$ are presented in table \ref{tab:charged}. Since the coefficient $B_{\rho\rho}$ exceeds significantly the others, the parameter $\epsilon_\rho$ should be small in comparison with $\epsilon_{\sigma,a}$. The possible reason is that the mixing of different isospin states in vector channel could occur only through three-gluon annihilation. If one neglects $\rho$-meson contribution, the branching fraction of $\eta'\to\pi^{0}\pi^{+}\pi^{-}$ decay is
\begin{eqnarray*}
BR_{th}\left(\eta'\to\pi^{0}\pi^{+}\pi^{-}\right) & = & (22\pm3)\times10^{-4}
\end{eqnarray*}
for $\epsilon_{\sigma}=\epsilon_{\epsilon}=0.016\pm0.003$ and
\begin{eqnarray*}
BR_{th}\left(\eta'\to\pi^{0}\pi^{+}\pi^{-}\right) & = & (12\pm2)\times10^{-4}
\end{eqnarray*}
for $\epsilon_{\sigma}=0.022\pm0.04$, $\epsilon_{a}=-0.011\pm0.002$. In the first case the theoretical prediction for the branching fraction agrees with the experimental value, while in the second case a large discrepancy is observed. This discrepancy can be removed fixing the isospin violation parameter in $\rho$-meson channel. According to our analysis, experimental value of the branching fraction is reproduced by setting $\epsilon_{\rho}=(1.6\pm0.8)\times10^{-3}$ in the $\epsilon_{\sigma}=\epsilon_{a}=0.016\pm0.1$ and $\epsilon_{\rho}=(2.2\pm0.2)\times10^{-3}$ for $\epsilon_{\sigma}=0.022$, $\epsilon_{a}=-0.011$ parametrization. Just as it was expected, in both cases $\epsilon_{\rho}$ is about an order of magnitude smaller than the parameter $\epsilon_\sigma$. One can determine what of the variants is realized in nature only from the analysis of the dalitz-plot shape. Unfortunately, there is no experimental information about it yet.

\begin{figure} 
\includegraphics{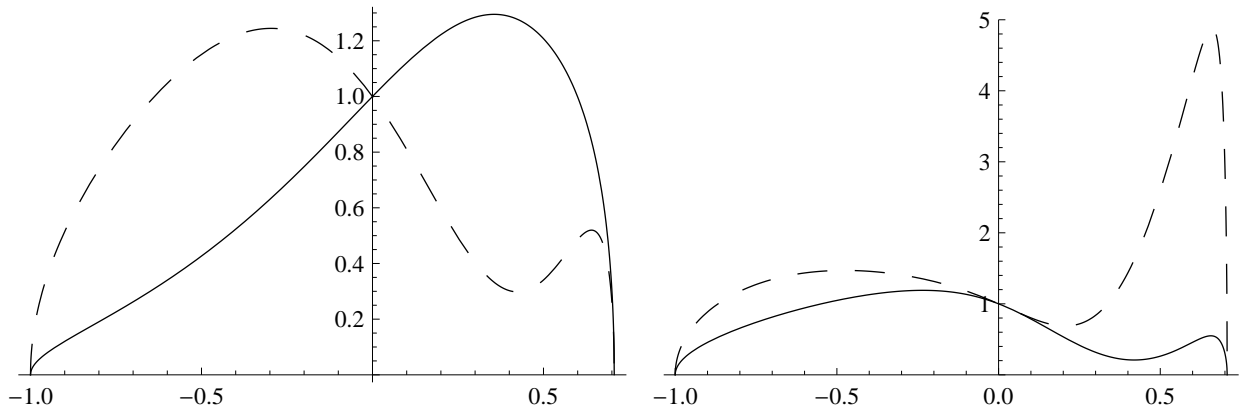}
\caption{
$Y$-distributions of squared matrix element of the decay $\eta'\to\pi^{0}\pi^{+}\pi^{-}$. Solid and dashed line correspond to parameter values $\epsilon_{\sigma}=\epsilon_{a}=0.016$ and $\epsilon_{\sigma}=0.022$, $\epsilon_{a}=-0.011$ respectively. On the left figure $\rho$-meson contribution is neglected. On the right figure the diagram with virtual $\rho$-meson exchange is taken into account. (the isospin violation parameter is equal to $\epsilon_{\rho}=1.6\times10^{-3}$ for the solid curve and $\epsilon_{\rho}=2.2\times10^{-3}$ for the dashed one}
\label{fig:Ydist2}
\end{figure}

The squared matrix element of this decay is usually written in the form of expansion
\begin{eqnarray}
\left|\mathcal{A}\right|^{2} & \sim & 1+aY+bY^{2}+cX+dX^{2}+\dots,
\label{eq:matr2}
\end{eqnarray}
where dalitz variables $X$, $Y$ are defined according to
\begin{eqnarray*}
X & = & \sqrt{3}\frac{T_{+}-T_{-}}{Q},\qquad Y=3\frac{T_{0}-m_{\pi^{0}}}{Q},
\qquad Q=M_{\eta'}-2m_{\pi^{+}}-m_{\pi^{0}}.
\end{eqnarray*}
In the above expression $T_{0}$ and $T_{\pm}$ are the energies of neutral and charged $\pi$-mesons in the initial meson rest frame. In fig.~\ref{fig:Ydist2} we show $Y$-distribution of the squared matrix element for different parametrizations. Sold and dashed curves on this figure correspond to parameter sets $\epsilon_{\sigma}=\epsilon_{a}=0.016$ and $\epsilon_{\sigma}=0.022$, $\epsilon_{a}=-0.0011$ respectively. In figure ~\ref{fig:Ydist2}a the contribution of $\rho$-meson was neglected, while in fig.~\ref{fig:Ydist2}b it was taken into account. It can be easily seen, that, in contrast to $\eta'\to3\pi^0$-decay, different parametrizations of the matrix element give significantly different distributions over the dalitz plot. In addition, despite of the smallness of isospin violation parameter in $\rho$-meson channel, the presence of this subprocess changes the distribution dramatically. For this reason experimental information on this distribution will be very useful.

\section{Conclusion}

The article is devoted to the investigation of dalitz distributions of isospin violating decays $\eta'\to3\pi^{0}$, $\pi^{0}\pi^{+}\pi^{-}$.

This decays are usually studied in the framework of chiral perturbation theory. It is well known, however, that for high energy reactions (for example, $\eta'\to 3\pi$) this theory is not valid. As a result, chiral perturbation theory does not describe well branching fractions and dalitz distributions for these decays \cite{Bijnens2007pr}.

We think that there is another, more physically motivated, approach for dalitz-plot description. In the framework of so called isobar model the matrix element of the considered process is saturated by contributions of virtual resonances with suitable masses and quantum numbers. In our recent paper \cite{Donskov2009ri} we use isobar model to study $\eta'\to\eta\pi^0\pi^0$ decay. It is shown in this work, that dalitz plot of this decay can be reproduced by taking into account two scalar resonances: $a_0$-meson in $\pi\eta$-channel and $\sigma$-meson in $\pi\pi$-channel. The integrated width of $\eta'\to\eta\pi\pi$ decay is determined mainly by $a_0$-meson contribution, and for describing the shape of the dalitz-plot it is necessary also to take into account $\sigma$-meson contribution.

In our present paper we use the same approach to study $\eta'\to3\pi$ decays. It is well known, that in these decays $G$-parity is violated, so its thorough investigation could give additional information on the nature of this effect. Actually, in our model there are several scales of isospin violation: either in the $\eta'\to\sigma\pi$ decay vertex or in final $\pi$-meson interaction (i.e. in $a_0\to\pi\pi$ reaction). The role of each of these mechanism can be determined from the analysis of $\eta'\to3\pi^0$ decay dalitz-plot. Presented in our paper results show, that coupling constant of $\eta'\to\sigma\pi$ vertex is suppressed in comparison with isospin allowed $\eta'\to\sigma\eta$ coupling constant by a small factor of the order $(m_{d}-m_{u})/m_{s}\approx0.02$. This value agrees well with previous theoretical estimates. As for isospin violation in virtual $a_0$-meson decay, the suppression in this case is much stronger. This result is rather surprising and can be caused by large experimental uncertainty in the shape of the dalitz-plot. Therefore we studied also another variant, when isospin violation parameters are equal to each other and agree well with theoretical estimates. It is interesting to note, that dalitz-distributions in these cases are rather close to each other. We think that this question deserves more thorough experimental investigation.

In the last part of our paper we consider the charged mode of the $\eta'\to3\pi$-decay, i.e. the reaction $\eta'\to\pi^0\pi^+\pi^-$. In this case, in contrast to neutral decay mode, $\rho$-meson exchange can also give contribution into the matrix element. The isospin violation parameter in this channel turns out to be about an order of magnitude smaller ($\sim 10^{-3}$). The possible reason is that the mixing of different isospin states in vector channel can occur only via three-gluon annihilation. Extremely interesting is the dalitz-plot shape in this decay. It turns out, that, contrary to $\eta'\to3\pi^0$ reaction, variation of model parameters changes the form of this distribution significantly. The experimental information in this distribution could be very important.

\begin{acknowledgments}
This work was done under financial support of Russian Foundation for Basic Research (grants \#10-02-00191-a, 09-02-00132-a, 07-02-00417-a). The work of one of the authors (L.A.V.) was also supported by President grant (\#MK-110.2008.2), grant of Russian Science Support Foundation and noncommercial
foundation ”Dynasty”.
\end{acknowledgments}


\appendix

\section*{Amplitudes}

\subsection{$\mathcal{A}_{\sigma}$}

During last years chiral pertubation theory and Roy equations \cite{Caprini2005zr} make possible to desribe accurately the amplitude of $\pi\pi$-meson rescattering at low energies. Unitarity relations put rather strong constrains on this amplitude. First of all, for $s \le (2m_\pi)^2$ it should be real, while for $s\ge(2m_\pi)^2$ up to $KK$ prodution theshold its imaginary part should satisfy the relation
\begin{eqnarray*}
\mbox{Im}\left(\frac{1}{\mathcal{A}(s)}\right) & \sim & \sqrt{1-\frac{4m_{\pi}^{2}}{s}}.
\end{eqnarray*}
Moreover, according to chiral perturbation theory, this amplitude should be equal to zero at $s=s_{A}=m_{\pi}^{2}/2$ (so called Adler effect).

The amplitude of $\pi\pi$-rescattering, that satisfy listed above conditions can be expressed through the series in a variable
\begin{eqnarray*}
w(s) & = & \frac{\sqrt{s}-\sqrt{4m_{K}^{2}-s}}{\sqrt{s}+\sqrt{4m_{K}^{2}-s}},
\end{eqnarray*}
that transforms $s$-variable plane with cuts $s\le0$ and $s\ge4m_{K}^{2}$ into a unit disc $|w|le 1$. The itroduction of ap new variable improves the convergence of the series in the considered variable domain. The amplitude of $\pi\pi$-scattering is written in the form \cite{Yndurain2007qm}:
\begin{eqnarray}
\kappa t_{0}^{0}(s) & = & \kappa\left\{ \frac{m_{\pi}^{2}}{s-s_{A}}\left[\frac{2s_{A}}{m_{\pi}\sqrt{s}}+B_{0}+B_{1}w(s)+\dots\right]-i\sqrt{1-\frac{4m_{\pi}^{2}}{2}}\right\} ^{-1},\label{eq:t00}
\end{eqnarray}
where unkonwn constant $\kappa$ can be determined from the fit of experimental data.

The analysis of NA48/2 results in $K_{e4}$ decay \cite{Caprini2008fc,Batley2007zz} tells us that in the presented above series one can leave only first two terms with the coefficients
о первые два члена с коэффициентами
\begin{eqnarray}
B_{0} & = & 7.4,\qquad B_{1}=-15.1.\label{eq:B}
\end{eqnarray}
With these values of parameters the amplitude (\ref{eq:t00}) has the pole at $\sqrt{s}=(459+209i)$~MeV, that can be interpreted as a pole of $\sigma$-meson with mass and width equal to
\begin{eqnarray*}
M_{\sigma} & = & 459\,\mbox{MeV},\qquad\Gamma_{\sigma}=518\,\mbox{MeV}.
\end{eqnarray*}
It is clear, that there are other resonances (for example $f_0$-meson) in this channel. Since the parametrization (\ref{eq:t00}) was obtained from the fit of experimental data, it includes all these contributions automatically.

The numerical value of the $\kappa$ constant depends on the quantum numbers of the initial particle and can be determined from expeimental data. The analisys of the $\eta'\to\eta\pi^{0}\pi^{0}$ decay, presented in \cite{Donskov2009ri}, shows, that in $\eta'\to\sigma\eta\to\pi\pi\eta$ reaction reasonable agreement with experimental dalitz-plot is obseved for
\begin{eqnarray}
\kappa & = & -4.0.\label{eq:kappa}\end{eqnarray}
In our case there is $G$-parity violation in $\eta'\to\sigma\pi$, that can be taken into account by small suppression factor.

Recalling all mentioned above we use the followinng form of the amplitude of $\sigma$-meson subprocesses:
\begin{eqnarray}
\epsilon_{\sigma}\mathcal{A}_{\sigma} & = & \epsilon_{\sigma}\kappa\left[t_{0}^{0}(s_{12})+t_{0}^{0}(s_{13})+t_{0}^{0}(s_{23})\right],\label{eq:AmpSigma}
\end{eqnarray}

Here $\epsilon_{\sigma}$ is a small suppression factor of the order $(m_{d}-m_{u})/m_{s}$, and $s_{ij}=(p_{\pi i}+p_{\pi j})^{2}$ are invariant masses of pairs of final pariticles. The following relation holds for these variables:
 \begin{eqnarray}
s_{12}+s_{13}+s_{23} & = & M_{\eta'}^{2}+3m_{\pi}^{2}.\label{eq:sum}
\end{eqnarray}

\subsection{$\mathcal{A}_{a}$}

It is well known, that the mass of $a_0$ meson is close to $KK$ production threshold. As a result, that propagator of this meson has a characteristic form different from usual Breit-Wigner parametrization. One has to introduce the energy-dependend self-energy part, caused by $\pi\eta$, $\pi\eta'$ and $KK$ loops, into the propagator. It is clear, however, that these corrections are significant only near $a_0$-meson pole. For the allowed in $\eta'\to3\pi$-decay mass region they are not very important. More important is the form of $a_0\eta'\pi$ and $a_0\pi\pi$ interaction vertices.

There are several ways to parametrise these verticies. One can use simple point-like interaction with coupling constants determined from experiment. Our analisys of $\eta'\to\eta\pi^{0}\pi^{0}$-decay shows, however, that these constants do not satisfy the relation of $SU(3)$-symmetry \cite{Feldmann1998sh}. The other parametrization is connected with chiral perturbation theory, that requres the interaction amplitude to be equal to zero in $p_\pi\to0$ limit. In the work \cite{Donskov2009ri} it was shown, that if one uses expressions $\gamma_{\pi\eta}(p_{\pi}p_{\eta})$ and  $\gamma_{\pi\eta'}(p_{\pi}p_{\eta'})$ for $a_{0}\pi\eta$ and  $a_{0}\pi\eta'$ interaction verticies, then the fit of the experimental dalitz-plot gives comparable values of $\gamma$-constants, as it is required by $SU(3)$-symmetry:
\begin{eqnarray}
\gamma_{\pi\eta} & \sim & \gamma_{\pi\eta'}\sim7\,\mbox{GeV}^{-1},\qquad\gamma_{\pi\eta}\gamma_{\pi\eta'}=35\,\mbox{GeV}^{-2}.
\label{eq:gamma}\end{eqnarray}
In our present work we use the same parametrization. The only difference is that there is isospin violation in $a_0\to\pi\pi$ decay. This effect can be parametrized introducing the small suppression factor of the order $(m_{d}-m_{u})/m_{s}$. So, we use the following expression for the $a_0$-meson exchange amplitude:
\begin{eqnarray}
\epsilon_{a}\mathcal{A}_{a} & = & \epsilon_{a}\gamma_{\pi\eta}\gamma_{\pi\eta'}\frac{\left(p_{\pi1}p_{\pi2}\right)\left(p_{\eta'}p_{\pi3}\right)}{s_{12}-m_{a}^{2}+im_{a}\Gamma_{a}\left(s_{12}\right)}+\mbox{premitations},\label{eq:AmpA0}\end{eqnarray}
Here the width of $a_0$-meson is determined by corresponding loop integral and $\epsilon_a$ is a small parameter that will be determined from the fit of experimental data.
 
\subsection{$\mathcal{A}_{\rho}$}

In the case of charged decay mode there are diagrams with virtual $\rho^{\pm}$ exchange (neutral $\rho^0$-meson is forbidden by charged parity conservation). The vertex of $\rho^{\pm}$-meson decay into $\pi^{\pm}\pi^0$ pair can be written in the form $g_{\rho}\left(p_{\pi^{+}}-p_{\pi^{0}}\right)_{\mu}\varepsilon^{\mu}$, where $\varepsilon^{\mu}$ is the polarization vector of $\rho$-meson and $g_{\rho}$ is effective decay constant. Numerical value of this constant can be determined from $\rho$-meson decay width:
\begin{eqnarray*}
\Gamma\left(\rho^{\pm}\to\pi^{\pm}\pi^{0}\right) & = & \frac{g_{\rho}^{2}}{24\pi}\left(1-\frac{4m_{\pi}^{2}}{m_{\rho^{2}}}\right)^{3/2}m_{\rho}.
\end{eqnarray*}
Using tabular values of $\rho$-meson mass and width it is easy to obtain $g_{\rho}\approx4.2$ for this constant. The vertex of $\eta'\to\rho^{\pm}\pi^{\mp}$ can be written in the similar form, but the corresponding coupling constant should be suppresed by a small isospin violation factor. As a result, the amplitude of $\eta'\to\pi^{-}\pi^{+}\pi^{-}$ decay with virtual $\rho$-meson has the form
\begin{eqnarray*}
\epsilon_{\rho}\mathcal{A}_{\rho}\left(s_{13},s_{13}\right) & = & 2g_{\rho}^{2}\left[\frac{s_{12}-s_{23}}{s_{13}-m_{\rho}^{2}+im_{\rho}\Gamma_{\rho}}+\frac{s_{12}-s_{13}}{s_{23}-m_{\rho}^{2}+im_{\rho}\Gamma_{\rho}}\right].\end{eqnarray*}

\end{document}